\documentclass[10pt,a4paper,onecolumn]{article}
\usepackage[a4paper, total={6in, 8in}]{geometry}
\usepackage{setspace,url}
\usepackage{lineno,hyperref,multirow}

\usepackage{epsfig,amssymb,amsmath}
\usepackage[utf8]{inputenc}
\usepackage{amsmath,graphicx}
\usepackage{amssymb,amsmath,epsfig,url,mathrsfs} \usepackage{algorithm,algorithmic}
\usepackage[usenames, dvipsnames]{color}
\usepackage[framemethod=TikZ]{mdframed}
\usepackage{xcolor}
\usepackage{colortbl}
\usepackage[colorinlistoftodos]{todonotes}
\usepackage{enumitem}

\usepackage{titling}
\usepackage{graphicx}
\usepackage{subfig}

\usepackage[style=ieee]{biblatex}
\usepackage{amsthm}

\makeatletter
\newcommand{\mathleft}{\@fleqntrue\@mathmargin0pt}
\newcommand{\mathcenter}{\@fleqnfalse}
\makeatother
 
\theoremstyle{definition}

\hypersetup{
   unicode=false,          
   pdftoolbar=true,        
   pdfmenubar=true,        
   pdffitwindow=false,     
   pdfstartview={FitH},    
   pdftitle={ASVspoof 2021 Evaluation Plan},    
   pdfauthor={ASVspoof},     
   pdfsubject={ASVspoof 2021},   
   pdfcreator={},   
   pdfproducer={}, 
   pdfkeywords={ASVspoof, automatic speaker verification and countermeasures challenge, spoofing, presentation attack detection}, 
   pdfnewwindow=true,      
   colorlinks=true,       
   linkcolor=blue,          
   citecolor=blue,        
   filecolor=magenta,      
   urlcolor=blue           
}

\mdfdefinestyle{MyFrame}{%
    linecolor=black,
    outerlinewidth=.1pt,
    roundcorner=1pt,
    innertopmargin=\baselineskip,
    innerbottommargin=\baselineskip,
    innerrightmargin=10pt,
    innerleftmargin=-10pt,
    backgroundcolor=green!10!white}

\title{\textbf{ASVspoof 2021:\\ Automatic Speaker Verification Spoofing and Countermeasures Challenge Evaluation Plan}\thanks{\textcolor{blue}{Document Version 0.4 (July 16, 2021)}}}

\author{
  Héctor Delgado, Nicholas Evans, Tomi Kinnunen, Kong Aik Lee,\\ Xuechen Liu, Andreas Nautsch, Jose Patino, Md Sahidullah,\\ Massimiliano Todisco, Xin Wang, Junichi Yamagishi\\\\
  ASVspoof consortium\\\url{http://www.asvspoof.org/}
}
\date{July 16, 2021}

\begin{document}

\maketitle



\begin{mdframed}[style=MyFrame]
\center{\underline{\textbf{TL;DR for new participants}}}
\begin{itemize}
\setlength\itemsep{0.25em}
    \item Your task is to develop a \emph{bona fide - spoofed} classifier (spoofing countermeasure) for speech data.
    \item We provide labelled training/development data and baseline code. You apply your system to unlabelled test data (one score per audio file) and send your scores to us, along with a system description.
    \item We rank and analyse the results, and present a summary at an INTERSPEECH 2021 satellite workshop. You are encouraged to submit a paper.
\end{itemize}
%
\center{\underline{\textbf{... and for readers familiar with ASVspoof}}}
\begin{itemize}
\setlength\itemsep{0.25em}
    \item ASVspoof 2021 involves technically more challenging data to promote countermeasure generalisation:
        \begin{itemize}
            \item A {\bf logical access} (LA) task similar to those of ASVspoof 2015 and 2019, but involving the \textbf{coding and transmission} of text-to-speech (TTS) and voice conversion (VC) attacks;
            \item A {\bf physical access} (PA) task involving \textbf{real} bona fide and replayed samples similar to ASVspoof 2017, but with a \textbf{better controlled design} (similar to the ASVspoof 2019 PA task).
        \end{itemize}    
    \item \textbf{A new speech deepfake task} (DF), involving compressed audio similar to the LA task (but without ASV).
    \item \textbf{Protocols:} Training/development partitions are the same as for ASVspoof 2019.  The evaluation partition will be new. 
    \item \textbf{Metrics}: for the LA and PA tasks, a slightly revised t-DCF  
    metric~\cite{Kinnunen-tDCF2020}. For the DF task, an equal error rate (EER) metric. 
\end{itemize}
\end{mdframed}

\section{Introduction}


The automatic speaker verification spoofing and countermeasures (ASVspoof) challenge series is a community-led initiative which aims to promote the consideration of spoofing and the development of countermeasures.  ASVspoof 2021 is the 4th in a series of bi-annual, competitive challenges where the goal is to develop countermeasures capable of discriminating between bona fide and spoofed or deepfake speech. 

ASVspoof 2021 comprises three different tasks:

\begin{itemize}
    
    \item logical access (LA): bona fide and spoofed utterances generated using text-to-speech (TTS) and voice conversion (VC) algorithms are communicated across telephony and VoIP networks with various coding and transmission effects;
    
    \item physical access (PA): bona fide utterances are made in a real, physical space in which spoofing attacks are captured and then replayed within the same physical space using replay devices of varying quality;
    
    \item speech deepfake (DF): a fake audio detection task comprising bona fide and spoofed utterances generated using TTS and VC algorithms. Similar to the LA task (includes compressed data) but without speaker verification. 
    
\end{itemize}

The LA and PA tasks maintain the focus of ASVspoof upon automatic speaker verification (ASV).  Accordingly, the metric for these two tasks will be the 
minimum tandem decision cost function (min t-DCF)~\cite{Kinnunen-tDCF2020}. 
The new DF task has a fake media / fake audio / deepfake flavour in which there is no ASV system.  The metric for the DF condition will revert to the equal error rate (EER).

This document provides a technical description of the ASVspoof 2021 challenge, including details of training, development and evaluation data, metrics, baselines, evaluation rules, submission procedures and the schedule.  It is intentionally terse.  Many aspects of the ASVspoof 2019 are identical to previous editions of ASVspoof.  The similarities are not repeated.  The focus in this document is instead upon the key details and differences to previous editions.

\section{Technical objectives}


While the promotion of \emph{generalised} countermeasures remains, the specific objectives for ASVspoof~2021 are centred around the shift towards more practical, realistic scenarios. Moving beyond the technical focus of previous editions, the objectives for the 2021 challenge are to: 

\begin{itemize}
\item promote the development of spoofing countermeasures that are robust to channel variability;
\item improve the reliability of spoofing countermeasures in less simulated, more realistic conditions, i.e. to recordings (re-)captured in real physical spaces; 
\item evaluate impacts of data augmentation to replace in-domain training and development data;
\item broaden the relevance of ASVspoof to non-ASV scenarios involving the detection of deepfakes.
\end{itemize}

\section{Logical Access (LA) task}

Given the relevance to telephony scenarios, the focus of ASVspoof 2021 for the LA task is upon the development of spoofing countermeasures that are robust to codec and transmission channel variability.  Bona fide and spoofed speech data generated with either text-to-speech (TTS), voice conversion (VC) or hybrid algorithms (voice conversion systems fed with synthetic speech) is transmitted across either a public switched telephone network (PSTN) or a voice over Internet Protocol (VoIP) network using some particular codec, prior to being processed by spoofing countermeasure and automatic speaker verification systems.  
Evaluation data will include trials transmitted across VoIP networks that use \texttt{alaw} and \texttt{G.722} codecs, though other codecs will also be used.
Speech data may hence exhibit coding and transmission artefacts as well as differences in bandwidths and sampling frequencies, in addition to artefacts related to spoofing, but no additive noise.  All data, no matter what the bandwidth, will be provided in a common 16~kHz, 16 bits per sample \texttt{FLAC} format.  Neither new training nor new development data will be distributed; participants will receive new evaluation data only.  Spoofing countermeasure systems should be trained and optimised using ASVspoof 2019 training and development data.  

The TTS, VC and hybrid spoofing attacks will be the same as those for the ASVspoof 2019 LA evaluation partition.  Codec meta data will not be provided for individual utterance.  
The task will hence be to design spoofing countermeasures that are capable of discriminating between bona fide and spoofed speech generated with known attacks when both contain additional nuisance variation relating to potential coding and transmission.  The primary metric will be the tandem detection cost function (t-DCF).

\section{Physical Access (PA) task}

\begin{figure}[t]
\centering
\includegraphics[trim=0 80 0 20, clip, width=\linewidth]{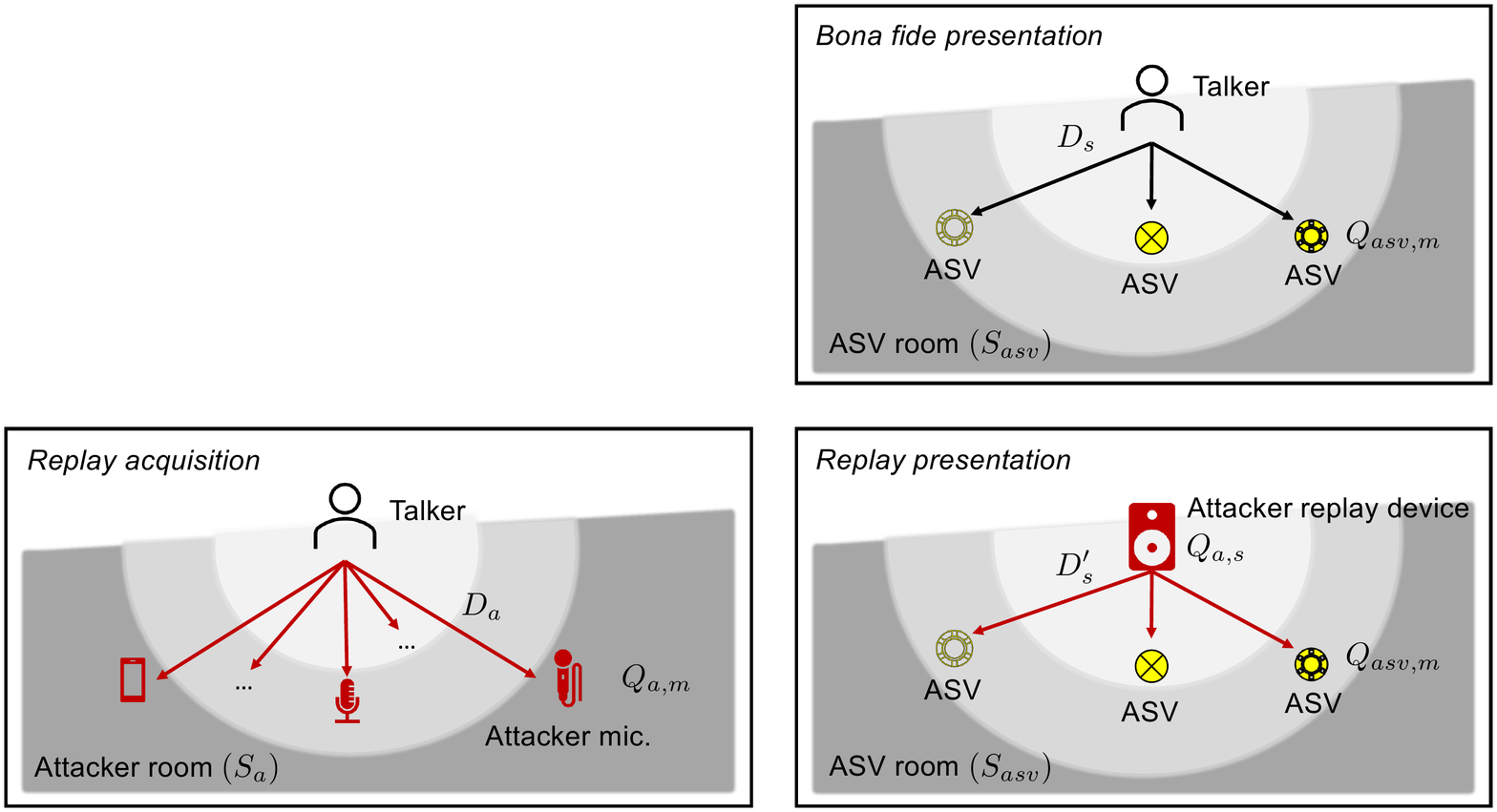}
\caption{An illustration of the ASVspoof 2021 physical access task. Rooms for replay acquisition (i.e., rooms in which the attacker captures recordings of the talker) and presentation (i.e., the room in which is situated the ASV system) may be different. In all cases the talker is simulated by a high quality loudspeaker. 
}
\label{fig:pa-illustration}
\end{figure}

The objective of the PA task is to develop countermeasures for the detection of replay spoofing attacks. In keeping with the ASVspoof 2019 challenge, participants should use the ASVspoof 2019 PA training and development data which contains simulated replay speech for countermeasure training and optimisation. Accordingly, training and development protocols for the PA task are identical to those for ASVspoof 2019. As for the LA task, no new training or development data will be released. 

New evaluation data will be distributed. This will contain predominantly \emph{real} replayed speech, in addition to a smaller proportion of \emph{simulated} replayed speech.  Participants must therefore make use of simulated training and development replay data to train and optimise countermeasures which perform reliably in the case of real replay data. Real replay speech in the evaluation set is influenced by several environmental and attacker factors similar to, but more comprehensive to those of the ASVspoof 2019 challenge data. 

The principal environmental factors include:
\begin{itemize}
    \item ASV room size $S_{asv}$ ($m^2$): $10 \leq S_{asv} \leq 65$
    \item Talker-to-ASV distance $D_s$ (cm): $50 \leq D_s \leq 200$
    \item ASV microphone $Q_{asv,m}$: MEMS and condenser microphones 
\end{itemize}
and the attacker factors include:
\begin{itemize}
    \item Attacker room size $S_a$ 
    \item Attacker-to-talker distance $D_a$
    \item Attacker microphone $Q_{a,m}$, 
    \item Replay device quality $Q_{a,s}$, and  
    \item Replay-device to ASV distance $D'_s$
\end{itemize}
These factors are illustrated in Figure~\ref{fig:pa-illustration}. The attacker factors are unknown. Neither meta information about the factors nor their labels will be provided. As a result of recordings being collected
in real physical spaces (rooms), 
all data contains low levels of background noise. 
Audio files will be delivered in the same \texttt{FLAC} file format as for the LA task.

\section{Speech deepfake (DF) task}

Similar to the LA task, the DF task involves discrimination of bona fide speech from artificially generated or converted speech that may have undergone compression. In contrast to the LA task, however, (1)~the DF task represents general audio compression (rather than telephony); and (2)~there is no ASV system. While no specific use case is implied, the task is potentially relevant to social media or forensics applications and situations where the aim of an adversary is to fool a human listener rather than an ASV system. 

The known compression algorithms include \texttt{mp3} and \texttt{m4a} while the evaluation data may also contain additional techniques. Here \emph{compression} refers to compression-decompression; the audio files will be delivered in the same \texttt{FLAC} file format as for other tasks, without meta data detailing the compression or processing applied. Since the aim is to promote the development of methods for the detection of speech deepfakes across multiple and unknown conditions, information neither about the compressor
(or presence/absense of compression) nor their settings (such as bitrate) will be provided. Similar to the LA task, the data originates primarily from the ASVspoof 2019 LA data. Additionally, there will be new data for the purpose of evaluating robustness to domain mismatch.

\section{Performance measures}

Performance measures estimate the Bayes risk, and thus reflect the expected cost resulting from using a countermeasure (CM) in product and service operation. For legacy only, we refer to the equal-error rate (EER) of a CM---we acknowledge that EER reporting is deprecated by ISO/IEC standards on performance testing in biometrics and presentation attack detection (CMs in our case)~\cite{ISO-IEC-197951-FDIS-2021,ISOpresentationAtack}. To estimate Bayes risk, one must quantify beliefs in class occurrence likelihoods (\emph{i.e.,} priors) and beliefs in costs resulting from erroneous class predictions (\emph{i.e.,} costs). The priors and costs shown in Table~\ref{tab:tDCF-parameters} are defined in the \href{sec:appendix}{Appendix} together with details of the min t-DCF computation. The ASV threshold is chosen to resemble the worst-case decision risk one can choose for the ASV subsystem, such that the optimisation task is specific to the CM subsystem (in an \emph{ASV-constrained} setting).

\begin{table*}[h]
\caption{t-DCF cost function parameters assumed in ASVspoof 2021.}
\label{tab:tDCF-parameters}
\begin{center}
\begin{tabular}{|c|ccc|ccc|}
\hline
							& \multicolumn{3}{c|}{\textbf{Priors}} & \multicolumn{3}{c|}{\textbf{Costs}}\\ 
Evaluation condition 		& $\pi_\text{tar}$ & $\pi_\text{non}$ & $\pi_\text{spoof}$ & $C_\text{miss}$ & $C_\text{fa}$ & $C_\text{fa,spoof}$\\
\hline\hline
LA task		& \multirow{2}{*}{0.9405} & \multirow{2}{*}{0.0095} & \multirow{2}{*}{0.0500} & \multirow{2}{*}{1} & \multirow{2}{*}{10} & \multirow{2}{*}{10}\\ 
PA task		& & & & & & \\  
\hline
\hline
DF task & \multicolumn{6}{c|}{Equal error rate (EER)} \\
\hline
\end{tabular}
\end{center}
\end{table*}

Note that, while these parameters are the same as for ASVspoof 2019, the ASV system and the conditions of the audio data are different. The above priors and costs, as well as ASV performance outline the ASV-constrained t-DCF computation. In consequence, the ASV-constrained t-DCF is computed according to three terms ($C_0, C_1, C_2$ -- see Appendix) in addition to the CM error rates. $C_0, C_1, C_2$ are different to those used for ASVspoof 2019, and will vary between 2021 progress and evaluation sets (see Section~\ref{sec:codalab}).

\section{Common ASV system and baseline countermeasures}

\subsection{Common ASV system}




The common ASV system is based on deep speaker embeddings followed by a PLDA-based scoring back-end. The speaker embedding extractor is trained on VoxCeleb data. The PLDA-based scoring back-end is trained on part of the same dataset and then adapted to each data condition. 
No ASV scores will be made available during the evaluation. 
Full details of the ASV system will be released at post-evaluation phase. 



\subsection{Baseline countermeasures}

A number of baseline CM solutions will be made available.
They include: a high resolution LFCC-GMM~\cite{sahidullah2015comparison,Tak2020} system; a CQCC-GMM~\cite{todisco2016new, todisco2017constant} system; an LFCC-LCNN~\cite{wang2021comparative} system; a RawNet2~\cite{9414234} system. Source codes for all systems are available from the following git repository:

\begin{center}
\url{https://github.com/asvspoof-challenge/2021}
\end{center}
Pre-trained models can be downloaded using scripts provided in the repository. 







\section{Evaluation rules}


The evaluation rules are common to each of the three tasks.  They are as follows.
\begin{itemize}
    \item Participants commit to using data from the ASVspoof 2019 countermeasures \emph{training} and \emph{development} partitions \textbf{only} to construct their countermeasures. The use of \emph{any} other external \emph{speech} dataset is strictly forbidden; this includes, but is not limited to, the use of any other public or in-house corpora, found speech or spoofed speech samples, externally trained models, feature vectors (or other statistical descriptors) extracted from external data, or externally trained speech activity detectors.  Use of ASVspoof 2019 \emph{evaluation} data as well as \emph{ASV} data is also forbidden.
    \item The use of external \emph{non-speech} resources (e.g.\ noise samples and impulse responses), speech codecs and audio compression tools for training or data augmentation \emph{is} permitted as long as their use is detailed in the system description accompanying any submission.
    \item Participants are allowed to re-partition the ASVspoof 2019 training and development data as they wish (e.g.\ by pooling the two to create a larger training set, or by excluding some training files).
    \item Each test sample must be scored independently; techniques such as domain adaption using the evaluation segments, or any use of  evaluation data for score normalisation is strictly forbidden.
    \item In contrast to ASVspoof 2019, participants may 
    submit scores for only \textbf{one} system
    and we do not require the submission of scores for development data. The challenge will be implemented using CodaLab. Participants will be permitted to update their scores multiple times during the progress phase.  During the evaluation phase, only a single submission will be permitted.  See Section~\ref{sec:codalab} for further details. 
    \item 
    Teams are required to submit a detailed system description relating to their 
    final submission (1 system description per team).  
    A system description template in the form of a high-level annotation will be provided and will serve as an indicator of the level of detail required.
    Compliance with the template and high-level annotation is mandatory  
    and is in everyone's interest.  System descriptions in such a common format will be essential to meaningful post-evaluation analysis and will help everyone  
    to put different approaches into context with one another (to promote open science and a better, common understanding).
    \item 
    Participants should not make direct, public comparisons to the results or rankings of competing teams.  This rule applies also to the disclosure of a team's own \emph{ranking} which is, by definition, a comparison to the results and rankings of competing teams.  A team can, however, publish their own \emph{results} in terms of t-DCF or EER and the like.  A summary of the challenge results will also be prepared by the organisers.  Any team wishing to retain anonymity should provide an anonymous team identifier with their registration.  The organisers reserve the right to adjust this policy prior to the scoring platform closing.
\end{itemize}

The organisers reserve the right to exclude systems from ranking and to exclude teams from future participation in ASVspoof in the event that the above rules are not adhered to.

\section{Registration and submission of results}


\vspace{-.5em}
\subsection{General mailing list}

All participants and team members are encouraged to subscribe to the general mailing list.  Subscribe by sending an email to:

\begin{center} \href{mailto:sympa@lists.asvspoof.org?subject=subscribe ASVspoof2021}{sympa@lists.asvspoof.org}
\end{center}

\noindent with `subscribe ASVspoof2021' as the subject line.  Successful subscriptions are confirmed by return email.  To post messages to the mailing list itself, emails should be addressed to:

\begin{center}
\href{mailto:asvspoof2021@lists.asvspoof.org?subject=ASVspoof 2021}{asvspoof2021@lists.asvspoof.org}
\end{center}

Subscribers can unsubscribe at any time by following the instructions in the subscription confirmation email.

\vspace{-.5em}
\subsection{Registration and database download}



Participants/teams are requested to register for the evaluation. Registration should be performed once only for each participating entity and by accessing the following registration form:

\begin{center}
    \url{https://www.spsc-sig.org/form/asvspoof-2021-registration-form}
\end{center}

The organisers aim to validate all registrations within 48 hours. Validated registrations will be confirmed by email.

There are three databases, one for each of the three tasks, all available from the Zenodo website:
\begin{center}
Logical Access:
\url{https://zenodo.org/record/4837263}
\end{center}

\begin{center}
Physical Access:
\url{https://zenodo.org/record/4834716}
\end{center}

\begin{center}
Speech Deepfake:
\url{https://zenodo.org/record/4835108}
\end{center}
Each dataset contains a list of evaluation data and corresponding audio files. All audio files are distributed in \texttt{FLAC} format. All audio data is sampled at a rate of 16 kHz and stored in 16-bit.

\vspace{-.5em}
\subsection{Submission of results}
\label{sec:codalab}

The ASVspoof 2021 challenge will be hosted on CodaLab:

\begin{center}
Logical Access:
\url{https://competitions.codalab.org/competitions/32343}
\end{center}

\begin{center}
Physical Access:
\url{https://competitions.codalab.org/competitions/32347}
\end{center}

\begin{center}
Speech Deepfake:
\url{https://competitions.codalab.org/competitions/32345}
\end{center}

In order to access each site, participants will be required to create an account on the CodaLab website.  This will enable you to request access to each submission site.  Teams should make only one request per task and requests should normally be made in the name of the team's registered contact person.  Requests will be validated manually by ASVspoof 2021 organisers.  The leaderboard is anonymous; usernames will not be disclosed to other participants.

The challenge will run in 2 phases, a \emph{progress} phase and the main \emph{evaluation} phase.  During the progress phase, each team may make up to 3 submissions per day. Results determined from a subset of trials will be made available for each submission.  During the evaluation phase, participants will be allowed only a single submission for which results will be determined from the remaining evaluation trials.  The schedule is shown in Section~\ref{sec:schedule}.

Results for 4 reproducible baselines (LFCC-GMM, CQCC-GMM, LFCC-LCNN, RawNet2) are displayed on the Codalab site for each task.

\subsection{System descriptions}

Participants must submit a system description for each task they participate via the following web forms:

\begin{center}
Logical Access:
\url{https://www.spsc-sig.org/form/asvspoof2021-la-systems}
\end{center}

\begin{center}
Physical Access:
\url{https://www.spsc-sig.org/form/asvspoof2021-pa-systems}
\end{center}

\begin{center}
Speech Deepfake:
\url{https://www.spsc-sig.org/form/asvspoof2021-df-systems}
\end{center}

\section{ASVspoof 2021 Workshop}


The \href{https://www.asvspoof.org/workshop}{ASVspoof 2021 Workshop} (an official \href{https://www.interspeech2021.org/satellites}{Interspeech 2021 satellite event}) is planned for September 16, 2021.  Due to uncertainty related to the global pandemic, the workshop will be organised fully online (details TBA). Participation will be free of charge and open to anyone (i.e.\ not limited to challenge participants). Challenge participants are encouraged to submit papers which describe their systems and results.  Besides challenge papers, we welcome research contributions related (but not limited) to speaker recognition, anti-spoofing, text-to-speech, voice conversion and speech deepfakes.
Submitted papers will be subject to blind peer review. Accepted papers will be published in the ISCA archive\footnote{\url{https://www.isca-speech.org/iscaweb/index.php/archive/online-archive}} and a DOI will be assigned to each paper. While the usual requirements regarding novelty are more relaxed for challenge system papers, reviewing criteria are independent to challenge results and ranking.  The usual requirements will apply to regular research contributions. 





\section{Schedule}
\label{sec:schedule}


\noindent A tentative schedule is as follows:

\begin{description}
\item \textbf{Challenge}
\begin{itemize}
\item Initial release of eval plan:\hspace*{\fill}April 19, 2021
\item Registration opens:\hspace*{\fill}April 19, 2021 
\item Evaluation data available:\hspace*{\fill}May 28, 2021
\item Registration deadline:\hspace*{\fill}June 23, 2021
\item Scoring platform closes:\hspace*{\fill}July 14, 2021 (AoE) 
\item System descriptions:\hspace*{\fill}July 16, 2021
\end{itemize}
\item \textbf{Workshop}
\begin{itemize}
\item Paper submission deadline:\hspace*{\fill}July 16, 2021
\item Notifications:\hspace*{\fill}August 16, 2021
\item Camera-ready papers:\hspace*{\fill}August 30, 2021

\item ASVspoof 2021 Workshop (virtual):\hspace*{\fill}September 16, 2021
\end{itemize}
\end{description}


\section{Glossary}

Generally, the terminologies of automatic speaker verification are consistent with that in the NIST speaker recognition evaluation. Terminologies more specific to spoofing and countermeasure assessment are listed as follows:

\textit{\textbf{Spoofing attack:}} An adversary, also named impostor, attempts to deceive an automatic speaker verification system by impersonating another enrolled user in order to manipulate speaker verification results.

\textit{\textbf{Anti-Spoofing:}} Also known as countermeasure. It is a technique to countering spoofing attacks to secure automatic speaker verification.

\textit{\textbf{Bona fide trial:}} A trial in which the speech signal is recorded from a live human being without any modification. 

\textit{\textbf{Spoof trial:}} In the case of the physical access, a spoofing trial means a trial in which an authentic human speech signal is first played back through an digital-to-analog conversion process and then re-recorded again through analog-to-digital channel; an example would be using smartphone $A$ to replay an authentic target speaker recording through the loudspeaker of $A$ to the microphone of smartphone $B$ that acts as the end-user terminal of an ASV system. In the case of the logical access, a spoofing trial means a trial in which the original, genuine speech signal is modified automatically in order to manipulate ASV.

\newpage


\printbibliography

\newpage

\section*{Appendix: evaluation metric details}
\label{sec:appendix}


The t-DCF \cite{Kinnunen-tDCF2020} is based on statistical detection theory and involves specification of an envisioned \emph{application}. A key feature of t-DCF is the assessment of a tandem system while keeping the two subsystems (CM and ASV) isolated from each other; 
the participant will develop standalone CM and submit their detection scores that the organizers will combine with the ASV scores. Thus, the participant does not need to develop, optimize or execute any ASV system at any stage. To enable rapid kick-off, the organizers provide t-DCF scoring code and provide pointers to existing CM implementations. Besides t-DCF, the organizers will evaluate equal error rate (EER) of each submitted countermeasure as a secondary metric. This section details the evaluation metrics. 

The ASVspoof 2021 challenge is to develop a two-class \emph{bona fide} vs.\ \emph{spoof} classifier. The positive (bona fide) class corresponds to speech segments of natural 
and genuine persons (combination of target and non-target classes) and the negative (spoof) class corresponds to attacks. The challenge participants are provided labeled audio files for development purposes (same as in 2019) while the evaluation set samples will be unlabeled. 

Both the t-DCF and EER metrics are computed from a set of \emph{detection scores}, one score corresponding to a single audio file (evaluation trial). Participants should assign a real-valued, finite numerical value to each trial which reflects the support for two competing hypotheses, namely that the trial is bona fide or spoofed audio. Similar to the former two edition of ASVspoof, \textbf{high detection score should indicate bona fide and low score should indicate spoofing attack}. The scores might be, for instance, log-likelihood ratios (LLRs) between bona fide vs. spoof hypothesis, bonafide posterior probabilities or inner product scores, to give examples. 



The primary metric of the LA and PA tasks is based on \emph{ASV-constrained} t-DCF \cite{Kinnunen-tDCF2020}, defined as
\begin{mdframed}[style=MyFrame]
    \begin{equation}\label{eq:tDCF-ASV-constrained}
        \text{t-DCF}(\tau_\text{cm}) =  C_0+C_1P_\text{miss}^\text{cm}(\tau_\text{cm})
        +C_2P_\text{fa}^\text{cm}(\tau_\text{cm}),
    \end{equation} 
\end{mdframed}    
where $P_\mathrm{miss}^\text{cm}(\tau_\text{cm})$ and $P_\mathrm{fa}^\text{cm}(\tau_\text{cm})$ are, respectively, the empirical miss and false alarm rates of the CM at threshold $\tau_\text{cm}$, computed by 
    \begin{eqnarray}
    P_\mathrm{miss}^\text{cm}(\tau_\text{cm})      & = & \frac{\#\{\mathrm{bona\; fide\;trials\;with\;CM\;score} \leq \tau_\text{cm}\}}{\#\{\mathrm{Total\;bona\;fide\;trials}\}}\nonumber\\
        P_\mathrm{fa}^\text{cm}(\tau_\text{cm})   & = & \frac{\#\{\mathrm{spoof\;trials\;with\;CM\;score} > \tau_\text{cm}\}}{\#\{\mathrm{Total\;spoof\;trials}\}},\nonumber
    \end{eqnarray}
and where $C_0$, $C_1$, and $C_2$ depend both on the t-DCF parameters and the ASV error rates.  They are given by:
    \begin{equation}\label{eq:tdcf-coefficients}
        \begin{aligned}
            C_0 & = \pi_\text{tar}C_\text{miss}P_\text{miss}^\text{asv}+\pi_\text{non}C_\text{fa}P_\text{fa}^\text{asv}\\
            C_1 & = \pi_\text{tar}C_\text{miss} - \left(\pi_\text{tar}C_\text{miss}P_\text{miss}^\text{asv}+\pi_\text{non}C_\text{fa}P_\text{fa}^\text{asv} \right) \\
            C_2 & = \pi_\text{spoof}C_\text{fa,spoof}
            P_\text{fa,spoof}^\text{asv},
       \end{aligned}
    \end{equation}
The meaning of the terms in Eq. \eqref{eq:tdcf-coefficients} are 
    \begin{itemize}
        \item $\pi_\text{tar},\pi_\text{non},\pi_\text{spoof}$ --- \textbf{a priori probabilities} of target user, non-target user, and a spoofing attack, respectively (nonnegative and sum = 1).
        \item $C_\text{miss}, C_\text{fa}, C_\text{fa, spoof}$ --- nonnegative \textbf{detection costs} assigned to missed target, falsely accepted nontarget and falsely accepted spoofing attack, respectively.
        \item $P_\text{miss}^\text{asv}, P_\text{fa}^\text{asv}, P_\text{fa,spoof}^\text{asv}$ --- \textbf{ASV system error rates} at a pre-specified ASV threshold --- miss rate of targets, false accept rate of nontargets, and false accept rate of spoofing attacks, respectively.
    \end{itemize}
    
The priors/costs shown in Table~\ref{tab:tDCF-parameters} do not to reflect some real-world case, which is subjective to and thus different for each participant, but---to reflect security-centred settings in which submitted systems are better to distinguish from one another.\footnote{
    To reflect priors and costs of, e.g., a specific business case for, particular services and products requires owners of these to quantify their own beliefs. This would be necessary in any case, even when CMs are compared by means of EER. When referring to EER, risk assessment is implied indirectly (Which EER is good? Why even equal-errors?): it is outsourced to owners of products and services by making lots of hidden assumptions. Here, we are fully transparent on our assumptions and provide formalism and reference implementations to Bayes risk reporting. When choosing more general/specific use cases for operation, priors/costs need to be adapted accordingly; expectations are contextual. 
} 
    
The raw t-DCF value can be difficult to interpret. Following the standard convention adopted in the NIST speaker recognition evaluation campaigns, it is useful to normalize the cost with that of a non-informative detector. In the realm of \eqref{eq:tDCF-ASV-constrained} this refers to a spoofing countermeasure which either accepts or rejects every input, whichever of these two yields lower t-DCF. A normalised version of the ASV-constrained t-DCF is therefore defined by
    \begin{equation}
        \text{t-DCF}(\tau_\text{cm})'=\frac{\text{t-DCF}(\tau_\text{cm})}{C_0 + \min\{C_1,C_2\}},
    \end{equation}
where the denominator is obtained by setting either $P_\text{miss}^\text{cm}(\tau_\text{cm})$ or $P_\text{fa}^\text{cm}(\tau_\text{cm})$ to 1 (in which case the other error rate is 0) and taking the minimum. Finally, as in 2019, we consider the ideally-calibrated countermeasure by adjusting the threshold $\tau_\text{cm}$ to the value whichever gives lowest $\text{t-DCF}(\tau_\text{cm})'$. That is, the primary metric is 
    \begin{equation}
        \text{t-DCF}_\text{min}' = \min_{\tau_\text{cm}}\,\text{t-DCF}(\tau_\text{cm})'.    
    \end{equation}
This metric ranges in between $[\texttt{ASV floor}, 1]$, where the value of 1 corresponds to a useless countermeasure (meaning the  system would have worked better without it) while the lower bound 
    \begin{equation}
        \texttt{Normalized ASV floor} \equiv \frac{C_0}{C_0 + \min\{C_1,C_2\}}
    \end{equation}
corresponds to a perfect (error-free) CM. We call it \emph{ASV floor} to indicate that the performance is floored by the ASV errors. 

The reader familiar with the DCF evaluation framework used in NIST speaker recognition evaluations (SREs) recognizes immediately the similarities\footnote{In fact, the NIST DCF \cite{NIST-SRE-CTSplan-2020} is a special case of the tandem t-DCF that assume zero spoof prior and which uses `accept all' countermeasure}. Nonetheless, unlike in the ASV tasks in the NIST SRE campaigns where the weightage of the miss and false alarms is \emph{known a priori}, note that the coefficients $C_0$, $C_1$ and $C_2$ depend not only on prior and cost terms but \emph{also} on the errors of the ASV system. The reader who participated ASVspoof 2019 recognizes that in ASVspof 2021 we retain the term $C_0$ in the metric (it was discarded in 2019). Note also that the cost formulation itself has been simplified from the original work \cite{Kinnunen2018-tDCF}; the reader is pointed to \cite{Kinnunen-tDCF2020} for further explanations. 

As in 2019, the challenge participant cannot interact with the ASV system and will have limited information about it. They \emph{will} be provided labeled ASV scores of the development set to demonstrate t-DCF computation, but will \emph{not} be provided any evaluation set ASV scores. The organizer's ASV system will be scored exactly on the same way on development and evaluation data but the evaluation set error rates may be different from those of the development set. 
What is common to all the ASV systems is that their detection threshold will be set at the equal error (EER) point, at which $P_\text{miss}^\text{asv}=P_\text{fa}^\text{asv}$ (see \cite[Section V]{Kinnunen-tDCF2020} for further discussion). 
Further information on ASV performance in one eval-set will be made available after the challenge.


Similar to the past two editions of ASVspoof, performance will be computed from detection scores pooled across all trials within a given challenge task (LA, PA or DF). That is, within one of these two tasks, all the bona fide trials in the evaluation set will be used to obtain $P_\text{miss}^\text{cm}(\tau_\text{cm})$ and all the spoof trials will be used to obtain $P_\text{fa}^\text{cm}(\tau_\text{cm})$
---ASV system error rates are determined by the challenge organisers.
This strategy is intended to encourage development of countermeasures that calibrate well across different variants of attacks and conditions. That is, to develop countermeasures whose scores are mutually compatible across a wide range of  conditions. The different challenge tasks, however, will be scored independent of each other. 
The three tasks should be considered independent tasks and a participant interested in all of them is welcome to use even completely different technologies to address each task.

The t-DCF parameters used in ASVspoof 2019 are provided in Table \ref{tab:tDCF-parameters}. The selected values, shared across the LA and PA tasks, are arbitrary but intended to be representative of \emph{authentication} applications. For instance, a banking domain may involve a very large number of authentication requests, most of which represent genuine target users; therefore, we have asserted high target prior. Priors of both `undesired' trials (nontargets and spoofing attacks) are set to low values for the same reason. The costs, in turn, reflect the idea that false acceptances are more detrimental than false rejections. Note that the priors do not necessarily reflect the empirical proportions of different classes in the evaluation data. 

\textbf{Secondary metric: equal error rate (EER):} The secondary evaluation metric is \emph{equal error rate} (EER) which was used in the past two editions of ASVspoof. EER corresponds to CM threshold $\tau_\mathrm{EER}$ at which the miss and false alarm rates equal each other\footnote{One may not find such threshold exactly as $P_\mathrm{fa}(\tau_\mathrm{EER})$ and $P_\mathrm{miss}(\tau_\mathrm{EER})$ change in discrete steps. You may use $\theta_\mathrm{EER}=\arg\min_\theta |P_\mathrm{fa}^\text{cm}(\tau_\text{cm})-P_\mathrm{miss}^\text{cm}(\tau_\text{cm})|$ or more advanced methods such as EER based on convex hull (ROCCH-EER) implemented in the Bosaris toolkit, \url{https://sites.google.com/site/bosaristoolkit/}. In ASVspoof2019, we use a simplified nearest-neighbor type computation demonstrated in the scoring package.}, i.e. $\mathrm{EER}=P_\mathrm{fa}^\text{cm}(\tau_\mathrm{EER})=P_\mathrm{miss}^\text{cm}(\tau_\mathrm{EER})$. 
Similar to t-DCF, EER will be computed from pooled evaluation trials within the given task.

\end{document}